# Quantification of Order in the Lennard-Jones System

by

Jeffrey R. Errington[¶§], Pablo G. Debenedetti[¶*], and Salvatore Torquato[‡]

¶ *Department of Chemical Engineering, Princeton University, Princeton, NJ 08544*

§ Current address: *Department of Chemical Engineering, State University of New York at Buffalo, NY 14260*

‡ *Department of Chemistry, Princeton University, Princeton, NJ 08544*

[*] Corresponding author. E-mail pdebene@princeton.edu

# Abstract

We conduct a numerical investigation of structural order in the shifted-force Lennard-Jones system by calculating metrics of translational and bond-orientational order along various paths in the phase diagram covering equilibrium solid, liquid, and vapor states. A series of non-equilibrium configurations generated through isochoric quenches, isothermal compressions, and energy minimizations are also considered. Simulation results are analyzed using an ordering map representation [Torquato et al., Phys. Rev. Lett. **84**, 2064 (2000); Truskett et al., Phys. Rev. E **62**, 993 (2000)] that assigns to both equilibrium and non-equilibrium states coordinates in an order metric plane. Our results show that bond-orientational order and translational order are not independent for simple spherically symmetric systems at equilibrium. We also demonstrate quantitatively that the Lennard-Jones and hard sphere systems sample the same configuration space at supercritical densities. Finally, we relate the structural order found in fast-quenched and minimum-energy configurations (inherent structures).

# I. Introduction

A problem commonly encountered by scientists and engineers is how to relate information found in a image of a material to its non-visual properties, for example its kinetic, mechanical, or electromagnetic properties. The image can come from a variety of sources, such as traditional imaging devices (e.g. various microscopes), as well as from numerical simulations, which generate representative configurations of a material. Examples of this type of relationship include the connection between the microstructure of a porous medium and the fluid flow characteristics through it, the microstructure of cheese and its textural characteristics [1], the organization of lipids in the skin and the rate of transdermal transport of drug molecules [2], and the cavity size distribution in bone and the onset of osteoporosis [3]. Although one can extract valuable information about a material by viewing an image of it, the visualization process is qualitative in nature, and thus results obtained from such a process will always possess a degree of subjectivity. To describe a material in a more objective manner, one must develop a formalism to describe quantitatively the information contained in images. One method for approaching this problem is to develop order metrics (order parameters) that identify given types of structural order within a system. Once these metrics have been formulated, they can be used to relate microscopic structural information to the macroscopic behavior of a system [4].

Numerous materials found in nature and utilized in industry exist in an amorphous state [5]. Such systems organize over relatively short distances, but lack the long-range order found in crystalline materials. Examples include liquids, window glass, tissue, various formulations found in the pharmaceutical industry, and numerous food products. Although methods for characterizing structural order in regular crystalline solids are well established [6,7], similar techniques for amorphous systems are not nearly as advanced. To quantify the structural order present in an amorphous system, one must first identify the types of order relevant to the system and subsequently construct metrics (preferably simple) that are capable of measuring that order. Recent studies involving the hard sphere system [8,9] and water [10] have made progress in this direction. In this work, we continue to develop these concepts by examining structural order in both equilibrium and non-equilibrium states of the Lennard-Jones system [11].

The reassessment of the notion of random close packing in hard-sphere systems by Torquato *et al.* [8] revealed a novel way of characterizing structural order. The authors defined metrics for two forms of structural order, bond-orientational and translational order, and introduced the concept of an ordering map, in which different states are mapped onto a plane whose coordinate axes represent the two order metrics. Subsequently, Truskett *et al.* [9] used the idea of an order map to identify the relative placement of a material's equilibrium phases in order metric space by focusing on hard-sphere systems. In addition, the ordering map can be used to trace the processing history of non-equilibrium structures. Once an ordering map has been constructed, it can be employed to interrogate relationships between a material's microstructure and macroscopic properties and/or to characterize complex material samples. For example, the ordering map was used recently to infer whether or not hard sphere configurations corresponded to thermally equilibrated states [9].

A detailed study of structural order in stable, supercooled and stretched liquid water has also been performed [10]. In water, directional attractions (hydrogen bonds) combine with short-range repulsions to determine the relative orientation of neighboring molecules as well as their instantaneous separation. The competition between these two interactions leads to the well-documented peculiar behavior of water [12]. By examining the relationship between structural order (orientational and translational) and the thermodynamic and kinetic properties of water, it was found that a cascade of anomalies occurs within the fluid, whereby structural, diffusive, and thermodynamic anomalies occur successively, as water becomes progressively ordered.

The objective of the present study is to examine structural order in a system of monatomic particles that interact in spherically symmetric fashion via soft repulsions and dispersive attractions. Specifically, we investigate the bond-orientational and translational order found in the vapor, liquid, and amorphous and crystalline solid phases of the shifted-force Lennard-Jones system. We first determine the phase boundaries for this system. We then examine the system along a number of paths involving the equilibrium vapor, liquid, and crystalline phases. We also consider non-equilibrium configurations generated through isochoric quenches, isothermal compressions, and energy minimizations.

The paper is organized as follows. In the following Section II we describe the order metrics employed in this work. The computational methods are detailed in Section III. In Section IV we discuss the results obtained in this investigation, and the salient conclusions are presented in Section V.

## II. Order Metrics

To quantify the structural order present in a material one must first identify a set of metrics that are sensitive to the types of particle arrangements relevant to the system. A natural method for determining the (dis)order in a system is to utilize a metric that quantifies the deviation of an actual structure from a reference arrangement, usually a crystalline lattice. Although this approach provides a relatively straightforward and accurate means of evaluating the (dis)order present in a material, the method has a number of limitations. The most serious one is the need to have knowledge of the structure of the reference crystalline phase. For simple systems, including the Lennard-Jones system considered here, this is generally not a problem. However, for more complicated condensed matter systems the crystalline structure may be unknown. Comparisons between the actual and reference structure, which are normally made at constant density [9], can also become ambiguous when multiple crystalline phases exist. In this case, the reference structure changes as a function of thermodynamic conditions. Furthermore, a stable solid phase will not exist at sufficiently low densities. For all of the above reasons, it is preferable to use crystal-independent metrics for the quantification of structural order. Our work is part of a broader study aimed at identifying suitable metrics for quantifying structural order in materials. To that end, we have focused here on order metrics that do not require knowledge of the stable crystalline structures. In what follows, we identify the types of order pertinent to our system and introduce the metrics used to quantify these forms of order. The reader is referred to Kansal *et al.* [13] for a discussion concerning the subtleties in choosing broadly applicable order metrics as well as guidelines for designing and evaluating new order metrics.

For a collection of spherically symmetric particles, there are two basic forms of order: bond-orientational order and translational order [8,9]. The first measures correlations between bond angles defined between a central particle and its nearest neighbors. The second measure quantifies the degree to which pairs of particles adopt preferential separations. To quantify the first, we employ a set of bond-orientational order metrics introduced by Steinhardt *et al.* [14]. The initial step in calculating the order metrics is to determine each particle's set of nearest neighbors. In this work, two particles are considered nearest neighbors if their separation is less than the radial distance to the first minimum in the pair correlation function. Following Steinhardt *et al.* [14], a vector $\mathbf{r}_{ij}$ pointing from a given molecule to one of its nearest neighbors is denoted as a "bond". For each bond one determines the quantity,

$$Q_{lm}(\hat{\mathbf{r}}_{ij}) = Y_{lm}(\theta_{ij}, \phi_{ij}), \tag{1}$$

where $\hat{\mathbf{r}}_{ij}$ is the unit vector of $\mathbf{r}_{ij}$ with the related polar and azimuthal angles $\theta_{ij}$ and $\phi_{ij}$, and the associated spherical harmonics $Y_{lm}$. Subsequently, an average over all bonds is performed to obtain, $\overline{Q}_{lm} = \langle Q_{lm}(\hat{\mathbf{r}}_{ij}) \rangle$. Finally, the averages $\overline{Q}_{lm}$, which depend on the choice of reference frame, are used to calculate the rotationally invariant order metrics $Q_l$.

$$Q_l = \left[ \frac{4\pi}{2l+1} \sum_{m=-l}^{m=l} |\overline{Q}_{lm}|^2 \right] \tag{2}$$

In this work, we have restricted our attention to even-$l$ spherical harmonics. In general, the order metrics grow in value as the crystallinity of a system increases. The limiting value of the order metric, for a perfect crystalline structure, depends on the value of $l$ and the type of crystalline lattice (see Figure 2 of reference [14]). For example, the $l = 6$ value for a perfect fcc crystal is $Q_6^{\mathrm{fcc}} = 0.57452$. For a completely uncorrelated system, the values of $Q_l$ become $1/\sqrt{N_{bond}}$, where $N_{bond}$ is the total number of bonds in the system. Therefore, in the infinite system limit the value of $Q_l$ spans from zero for a completely random system to the value for the perfect crystalline structure that it adopts.

To evaluate the translational order, we use a slight modification of the crystal-independent translational order metric introduced by Truskett *et al.* [9],

$$\tau = \frac{1}{s_c} \int_0^{s_c} |g(s) - 1| ds, \qquad (3)$$

where, $s = r\rho^{1/3}$ is the radial distance scaled by the number density, $g(s)$ is the pair correlation function, and $s_c$ is a numerical cutoff, which in this work was set to $s_c = 3.5$. The order metric provides a measure of the local density modulations over a finite number of coordination shells. For a completely uncorrelated system, $g(s) \equiv 1$, and thus $\tau$ has a value of zero. Conversely, the value of $\tau$ is relatively large for systems with long-range order. For a perfect fcc crystal, one can analytically determine the value of $\tau^{fcc}(s_c = 3.5) = 1.7893$.

## III. Simulations

A variety of numerical methods were employed in this work to determine the phase behavior and to examine the structural order in equilibrium and non-equilibrium configurations of the shifted-force Lennard-Jones (sfLJ) system. The specific form of the potential employed in this work is,

$$u^{sf}(r) = \begin{cases} u(r) - u(r_c) - (r - r_c)u'(r_c) & \text{for } r \leq r_c \\ 0 & \text{for } r > r_c \end{cases}, \qquad (4)$$

with,

$$u(r) = 4\varepsilon\left[\left(\frac{\sigma}{r}\right)^{12} - \left(\frac{\sigma}{r}\right)^{6}\right], \tag{5}$$

where $u^{\text{sf}}$ is the shifted-force potential energy, $u$ is the full Lennard-Jones interaction energy, $u'$ represents the first derivative of the full potential, $r_c$ is the radial distance of the potential cutoff ($r_c = 2.5$ in this work), and $\varepsilon$ and $\sigma$ are energy and size parameters, respectively. In what follows, all quantities are nondimensionalized using $\varepsilon$ and $\sigma$ as characteristic energy and length scales, respectively. For example, temperature is reduced by $\varepsilon/k_B$ ($k_B$ is the Boltzmann factor), distance by $\sigma$, and time by $\sqrt{m\sigma^2/\varepsilon}$, where $m$ is the mass of a particle, which is set to unity in this work.

Monte Carlo methods were used to determine the phase behavior of the system. The vapor-liquid phase boundary was determined using histogram reweighting grand canonical Monte Carlo [15]. The methods used here are analogous to those described elsewhere [16,17]. A series of grand canonical simulations were completed at state points in the vicinity of the coexistence curve. In particular, a run was completed at near critical conditions as well as six liquid phase and four vapor phase runs, with the lowest temperature simulated being $T = 0.57$. The histograms were combined using the techniques of Ferrenberg and Swendsen [18,19]. The volume of the simulation cell was set to $V = 216$.

The vapor-liquid critical point parameters were determined in a manner analogous to that employed by Potoff and Panagiotopoulos for the full LJ potential [20]. Specifically, a finite-size analysis [21,22] was used to calculate the apparent critical parameters for system sizes of $V = 343, 512, 1000$, and $1728$. Subsequently, this data was used to extrapolate to the infinite system size critical parameters.

The liquid-solid and vapor-solid phase equilibria were determined using the Gibbs-Duhem integration technique introduced by Kofke [23,24]. To implement this method, one needs to specify the type of crystalline lattice that the system adopts. In a recent study, Jackson *et al.* examined the relative stability of the fcc and hcp lattices for the full Lennard-Jones system [25]. They found that the fcc lattice was the stable phase for all pressures above a temperature of $T \approx 0.4$. Moreover, they determined that the liquid always freezes into the fcc lattice. However, the authors also showed that the relative stability of the fcc and hcp lattices can vary appreciably with changes in the method used to truncate the potential. We elected to perform the phase equilibrium calculations in this study assuming that the fcc lattice was the stable crystalline phase. Although for some of the crystalline state points considered in this work the fcc lattice may be metastable with respect to the hcp lattice, the conclusions from this work are independent of the precise nature of the stable crystalline lattice, especially since we quanitify structural order in crystal-independent fashion. The location of the liquid-crytalline phase transition would be altered by a negligible amount given the similarity in the free energy of the fcc and hcp lattices.

The methods used in this work are very similar to those employed by Agrawal and Kofke to determine the phase behavior of the soft sphere (SS), $u(r) = \varepsilon(\sigma/r)^n$ [26,27], and Lennard-Jones (LJ) systems [28]. The location of the liquid-solid equilibria for the sfLJ system was performed in three steps. First, the phase coexistence of the sfSS system (the SS potential truncated and shifted according to Equation (4)) was determined by integrating the change in the logarithm of the saturation pressure with potential softness, $\alpha \equiv 1/n$, from $\alpha = 0$ (hard sphere) to $\alpha = 0.085$. An integration step size of $\Delta\alpha = 0.005$ and a system size of $N = 500$ were used. Next, the attractive part, $u(r) = -4\varepsilon(\sigma/r)^6 \cdot \kappa$, of the sfLJ potential was added to a sfSS potential scaled by a factor of four, $u(r) = 4\varepsilon(\sigma/r)^{12}$, by integrating the variation of the logarithm of the saturation pressure with respect to $\kappa$ from $\kappa = 0$ to $\kappa = 1$ using a step size of $\Delta\kappa = 0.05$, while holding the temperature constant at $T = 2.74$. The remainder of the solid-liquid coexistence was then determined by integrating the change in the reciprocal temperature with the logarithm of the pressure in step sizes of $\Delta \ln p = -0.2$, using the final point from the previous integration as a starting point.

Before determining the vapor-solid equilibria, we first located the vapor-liquid-solid triple point. This was accomplished by identifying the point of intersection of the vapor-liquid and liquid-solid coexistence curves. The triple point then provided a starting point to obtain the sublimation line, which was determined by integrating the variation of the temperature with pressure in step sizes of $\Delta p = -0.0002$.

The structural order metrics were determined from data collected during $NVT$ molecular dynamics simulations. The equations of motion were integrated with the velocity-Verlet algorithm [29] coupled with a single Nosé-Hoover thermostat [30]. The time step was set to $\Delta t = 0.002$ and the Nosé-Hoover thermostat coupling constant $Q_{NH}$ was set to $Q_{NH} = 2.0$. A system size of $N = 500$ particles was used for all simulations. At low densities the initial fluid configurations were randomly generated and subsequently these configurations were rescaled for use as starting configurations for higher density runs. A perfect fcc crystal was used to initiate all simulations involving the crystalline phase.

Our study also included the examination of non-equilibrium structures. These structures were generated using three different methods: isochoric quenches, isothermal compressions, and isochoric energy minimizations. The isochoric quenches were completed by reducing the temperature at rates varying between $\Delta T/\Delta t = -0.005$ and $\Delta T/\Delta t = -0.2$ with step changes of $\Delta T = -0.1$. For typical argon parameters ($\varepsilon/k_B = 98.87$ K and $\sigma = 3.29$ Å) [31], this corresponds to quenches varying between $2.15 \cdot 10^{11}$ and $8.62 \cdot 10^{12}$ K/sec with step changes of 9.9 K. The quench was initiated by performing an equilibration run of length $t_e$, after which the final configuration was saved and a production run of length $t_p = t_e$ was completed. The temperature was then reduced by $\Delta T$ and the above process was repeated using the final configuration of the equilibration phase from the previous temperature as the starting configuration. This sequence of events was continued until a temperature of $T = 0.1$ was reached [32]. The isothermal compressions were performed in a manner analogous to the isochoric quenches. The compression rate was adjusted between $\Delta \rho/\Delta t = 0.005$ and $\Delta \rho/\Delta t = 0.1$ by increasing the density in increments of $\Delta \rho = 0.1$. The compression was terminated when a density of $\rho = 2.15$ was reached. This algorithm corresponds to argon compression rates varying between $4.06 \cdot 10^{12}$ and $8.12 \cdot 10^{13}$ kg/(m$^3$ sec) with steps of 186 kg/m$^3$.

Inherent structures [33] were the final type of non-equilibrium structure generated. These structures were generated by performing a conjugate gradient numerical approximation to steepest descent energy minimization of instantaneous equilibrium liquid configurations to the corresponding local potential energy minimum [34,35]. Each iteration of the conjugate gradient method causes a decrease in the potential energy. A minimization was considered complete when a new iteration yielded a relative change in the potential energy, $|(E_i - E_{i-1})/E_i|$, of less than $10^{-8}$. For each state point considered, at least 50 inherent structures were generated to determine the average value of properties. Although inherent structures are in mechanical equilibrium (minimum energy), they are not in thermal equilibrium, as their thermal energy has been removed. This justifies their being considered as non-equilibrium configurations.

## IV. Results and Discussion

We are not aware of any literature reference on the phase diagram of the shifted-force Lennard-Jones system. We therefore performed such calculations as a preliminary step in the present study. The phase diagram of the shifted-force Lennard-Jones system is presented in Figure 1 and the values of the critical and triple point parameters are collected in Table 1. As one would expect, the critical and triple point temperatures of the sfLJ potential are lower than those of the full LJ potential. However, while the liquid and solid triple point densities are lower for the sfLJ than in the LJ case, the critical density is very similar to that of the full LJ potential. Knowledge of the phase behavior is important in this study, as a number of the paths considered involve moving along phase boundaries.

Figure 2 shows the bond-orientational order metrics, $Q_l$, as a function of density along a supercritical isotherm of $T = 1.5$. The density range covered includes both the fluid and crystalline phases. The bond-orientational order metrics with $l = 4, 6, 8,$ and $12$ are sensitive to fcc ordering in the crystalline phase, with value increasing monotonically with density. The salient feature is the discontinuous jump in bond-orientational order at the fluid-crystal transition. Note that $Q_6$ is the most sensitive of the bond-orientational order metrics examined, since it produces the largest difference between the fluid and crystalline phases. As a result, $Q_6$ is used to quantify bond-orientational order in what follows. Although changes in the bond-order metrics with density in the fluid phase are much smaller than across the fluid-crystal transition, it is nevertheless worth noting that all of the bond-orientational order metrics initially decrease with density in the fluid phase. For $l$ = 6, 8, and 12 a minimum in the orientational order is observed at $\rho \approx 0.6$.

Figure 3 shows the behavior of $Q_6$ along the fluid and crystalline branches of the fluid-crystal transition, the vapor and liquid branches of the vapor-liquid transition, a subcritical isotherm at $T$ = 0.75, the critical isotherm, and a supercritical isotherm at $T$ = 1.5. In the crystalline phase, as one would expect, $Q_6$ increases with density at constant temperature and decreases with temperature at constant density. The value of $Q_6$ for the saturated crystal decreases monotonically with increasing temperature. Below a density of $\rho \approx 0.6$, $Q_6$ increases as the density decreases. This finding is in contrast to what was observed for the hard sphere fluid by Truskett *et al*. [8,9], where it was found that the value of $Q_6$ invariably increases with density. This difference in low-density behavior reflects the role of attractive interactions, which are present in the Lennard-Jones system. A detailed study is underway to investigate low-density configurations sampled as a result of attractive interactions.

The crystal-independent translational order metric, $\tau$, was calculated along the same paths on the phase diagram used for determining $Q_6$. The results of these calculations are presented in Figure 4. The observed behavior is qualitatively very similar to that of the bond-orientational order at higher densities. For both the fluid and crystalline phases, the value of $\tau$ increases with density at constant temperature and decreases with temperature at constant density. The translational order decreases with temperature along the crystalline branch of the fluid-crystal transition and remains relatively constant along the fluid branch. Along the vapor-liquid transition, translational order increases monotonically with saturated density. In the near-critical region the translational order metric remains relatively constant. This demonstrates that the translational order metric defined in Equation (3) is not sensitive to the long-range behavior of $g(r)$ caused by critical fluctuations. We also investigated the ability of the two-body excess entropy [36] to serve as a metric for translational order. In all cases qualitatively analogous results were found with this order metric.

The calculations of bond-orientational and translational order along different paths on the phase diagram allow the construction of an order map. This method of characterization, recently introduced by Torquato and coworkers [8,9], is a novel way of interrogating the structural order that exists in a given material. Figure 5 shows an order map with the translational order metric $\tau$ plotted versus the bond-orientational order metric $Q_6$. It can be seen that all of the data collected in the crystalline and fluid phases collapse onto single lines for each of the respective phases. This means that the two order metrics are strictly correlated for dense equilibrium states of the sfLJ system: translational and orientational order are not independently variable. While this is a reasonable result for a spherically symmetric system, the numerical results offer a clear confirmation.

It is interesting to compare the order map of the sfLJ (this work) and hard sphere [8,9] systems. This is done in Figure 6. It can be seen that the data for the hard sphere system and sfLJ system fall on the same lines. This suggests that identical types of order are generated by sufficiently dense simple spherically symmetric systems at equilibrium.

The same microscopic configuration is produced in these simple systems, with a change in state point or Hamiltonian resulting in a shift along the fluid or crystalline line on the order map. These concepts are not new: in essence, they form the basis of perturbation theory [37], which states that all simple fluids have the same underlying structure and that changing the details of the interaction can be treated as a small perturbation to a reference system. The results presented here provide precise and quantitative evidence to support the physical principle that underlies perturbation theory. They also yield a lower density limit ($\rho \approx 0.6$ for the sfLJ system) below which this picture breaks down because attractive interactions cause the system to sample configurations that are distinct from those accessed in a purely repulsive system at the same density. This can be seen by the pronounced *increase* in $Q_6$ as the density decreases along the saturated vapor branch (Figure 3).

The previous discussion pertained to equilibrium states. In what follows we extend the methodology to the investigation of supercooled liquid and glassy states. We consider systems generated by isochoric quenches, isothermal compressions, and energy minimizations of equilibrium liquid configurations. Figure 7 shows the paths traversed on the order map during isochoric quenches at various rates. The quenches were performed at a density of $\rho = 0.95$, starting from a temperature of $T = 1.5$ and terminating at $T = 0.1$. All of the quenches begin by uniformly extending the liquid branch on the order map to higher values of order. After this initial trend the trajectories separate, with the slower quenches attaining higher degrees of both bond-orientational and translational order. At the lower quench rates, sufficient time exists for crystallites to form, which significantly increase the overall structural order. Conversely, for fast quench rates, only a modest increase in structural order is observed. This inverse dependence of the quench rate on the order is consistent with the results for hard spheres [8,9], where compression rate plays a role analogous to quenching. Also included in Figure 7 is a collection of points representing the structural order found in inherent structures (local energy minimizations). These points were obtained from energy minimizations over a wide density range. In particular, equilibrium liquid configurations were generated at the relatively high temperature of $T = 50$ for densities ranging from $\rho = 0.9$ to 2.0, after which a number of the instantaneous configurations were subjected to energy minimization as described in Section III to obtain the inherent structures. The minimum density examined is just above the Sastry density for this system [35], which marks the lowest density for which the inherent structures are spatially homogeneous, and below which system-spanning cavities and fissures begin to form. All of the inherent structure data collapses onto a very narrow region on the order map. This suggests that minimum energy configurations correspond to rather specific particle arrangements that are quite insensitive to changes in density over the range of conditions studied here.

    The connection between the isochoric quench and inherent structure results has a number of interesting consequences. First, a reasonable extrapolation of the data indicates that a quench taken to zero temperature with a very high cooling rate leads to the same location on the order map as an energy minimization. This analysis implies that quenches performed at a rate that exceeds some system-specific high-cooling-rate limit are identical to energy minimizations, something that has long been suggested, but for which little quantitative evidence existed so far. Second, it appears that the collection of inherent structure points represents a smooth extrapolation of the liquid branch on the

order map, which indicates that inherent structures are the most ordered amorphous configurations attainable.

A series of isothermal compressions was performed in a manner similar to the isochoric quenches. Specifically, a system equilibrated at a temperature of $T = 1.5$ and a density of $\rho = 0.95$ was compressed linearly in density to $\rho = 2.15$. The data collected from these compressions are displayed in Figure 8. The results are qualitatively similar to the isochoric quenches. At low compression rates, a large number of crystallites form, causing the structural order to increase. As the compression rate is increased, fewer crystallites form, which results in final configurations with lower structural order. In contrast to the isochoric quenches, all of the trajectories generated from the compressions eventually veer away from the collection of inherent structures on the order map. The precise reason for this is unknown. One possible explanation is that in order to converge to the inherent structures, even higher compression rates are needed than those used in this work.

# V. Conclusions

In this study, we investigated the structural order found in equilibrium and non-equilibrium states of the shifted-force Lennard-Jones system. A number of accurate methods were used to determine the underlying phase behavior. Two forms of order, bond-orientational and translational, were then examined using simple metrics introduced recently by us [8,9]. Our results lead to a number of new insights into the structure of simple systems interacting via spherically symmetric forces, such as colloidal particles and rare gases.

The order map, which projects configurations onto a sub-space whose coordinate directions are the system's relevant order metrics, provides a useful representation of the evolution of structural order along both equilibrium and non-equilibrium paths. Our results show that all of the equilibrium data for the sfLJ system collapse onto two lines on the order map, one for the fluid and one for the crystalline phases, respectively. The corresponding data from the hard sphere system also fall on these same two lines. This analysis demonstrates that bond-orientational and translational order are not independent for simple spherically symmetric systems at equilibrium. Our work shows that simple systems are restricted to a well-defined line in bond-orientational – translational order space, and that a change in the state point or Hamiltonian simply causes a shift along that line. Stated more generally, the work quantitatively demonstrates that simple spherically-symmetric systems sample the same configurations. This relationship breaks down at sufficiently low densities, where attractive forces give rise to behavior not seen in hard-core systems.

Investigation of supercooled and glassy states resulted in several novel findings. The structural order generated by supercooled liquids upon isochoric quenches extends the fluid branch on the ordering map. For slower quenches, the formation of crystallites causes a sharp increase in structural order. At the fastest quench rates investigated here, the trajectories move toward the location of the inherent structure points on the ordering map. The data suggests that energy minimizations are a limit of very fast quenches and that inherent structures represent the natural extrapolation of the fluid branch on the ordering map.

Future work will focus on resolving open questions raised by this work, and on extending this approach to more complex systems. A detailed study of low-density behavior in systems with attractive forces (such as the sfLJ) is warranted. It is hoped that such an investigation will improve current understanding of the distinct types of particle arrangements brought about by dispersive forces at low densities, and their relationship to a system's thermodynamic and transport properties. Another open question involves structural order within the metastable vapor-liquid region. An important finding for the hard-sphere system [8,9] is the fact that non-equilibrium (glassy) states sample a distinct region of the order map; it is therefore of interest to explore what regions correspond to states that are metastable with respect to the vapor-liquid transition. It also appears important to investigate the extent to which dense spherically symmetric systems sample only hard-sphere configurations. To address this issue, we plan to examine the structural order found in systems with more complex spherically symmetric interactions, such as the Gaussian core model [38] and core softened systems [39,40]. It has recently been demonstrated that commonly employed order metrics may not be consistent with one another over a broad range of conditions [13]. Therefore, an important aspect of future research will involve the investigation and application of improved order metrics. Finally, we think that the quantification of structural order can also be used as a powerful tool for studying aqueous systems. Our recent work on water [10] represents a logical starting point for the analysis of aqueous solutions containing biological molecules, such as carbohydrates and amino acids, as well as ionic species. In these cases, orientational order metrics are likely to yield novel and quantitative insights into solvation phenomena and solution structure.

# Acknowledgements

We thank T. M. Truskett, V. K. Shen, and F. H. Stillinger for many valuable discussions and suggestions. PGD gratefully acknowledges the financial support of the Department of Energy, Division of Chemical Sciences, Geosciences, and Biosciences, Office of Basic Energy Sciences (grant DE-FG02-87ER13714).

# Tables

**Table 1.** Critical and triple point parameters of the shifted-force Lennard-Jones potential.

| Property | sfLJ | LJ |
|---|---|---|
| $T_c$ | 0.937 | 1.3120 |
| $P_c$ | 0.0820 | 0.1279 |
| $\rho_c$ | 0.320 | 0.316 |
| $T_{tr}$ | 0.560 | 0.687 |
| $P_{tr}$ | 0.0018 | 0.011 |
| $\rho_{tr,\,sol}$ | 0.936 | 0.963 |
| $\rho_{tr,\,liq}$ | 0.815 | 0.850 |
| $\rho_{tr,\,gas}$ | 0.00334 | 0.00186 |

# Figure Captions

**Figure 1.** Phase diagram of the shifted-force Lennard-Jones system in the temperature-density (a) and pressure-temperature (b) planes. The solid and dashed lines represent the vapor-liquid and liquid-solid phase boundaries respectively. The triangle, circle, and diamond indicate the vapor-liquid critical point and the saturated liquid and solid phases at the triple point.

**Figure 2.** Bond-orientational order metrics as a function of density at a temperature of $T = 1.5$. Symbols are as follows: (○) $Q_2$, (□) $Q_4$, (◇) $Q_6$, (△) $Q_8$, (◁) $Q_{10}$, (▽) $Q_{12}$.

**Figure 3.** Bond-orientational order metric $Q_6$ as a function of density along several loci on the phase diagram. Symbols are as follows: (—) vapor-liquid coexistence boundary, (– –) liquid-solid coexistence boundary, (△) vapor-liquid critical point, (○) liquid triple point, (◇) solid triple point, (□) $T = T_c = 0.935$ isotherm, (▷) $T = 0.75$ isotherm, (▽) $T = 1.5$ isotherm.

**Figure 4.** Translational order metric $\tau$ as a function of density along select paths on the phase diagram. Symbols are the same as in Figure 3.

**Figure 5.** Ordering map. Symbols are the same as in Figure 3.

**Figure 6.** Comparison of the Lennard-Jones (solid line) and hard sphere (circles) ordering diagrams.

**Figure 7.** Trajectories of isochoric quenches on the ordering map. The solid lines represent the equilibrium fluid and crystalline phases and the dashed lines correspond to quench rates $\Delta T/\Delta t = -0.005, -0.0067, -0.01, -0.02, -0.05, -0.1,$ and $-0.2$. The arrow indicates the direction of faster quenches. Diamond symbols indicate inherent structures.

**Figure 8.** Trajectories of isothermal compressions on the ordering map. Symbols are the same as in Figure 7. The dashed lines correspond to isothermal compression rates $\Delta\rho/\Delta t = 0.005, 0.01, 0.02, 0.05,$ and $0.1$. The arrow indicates the direction of faster compressions.

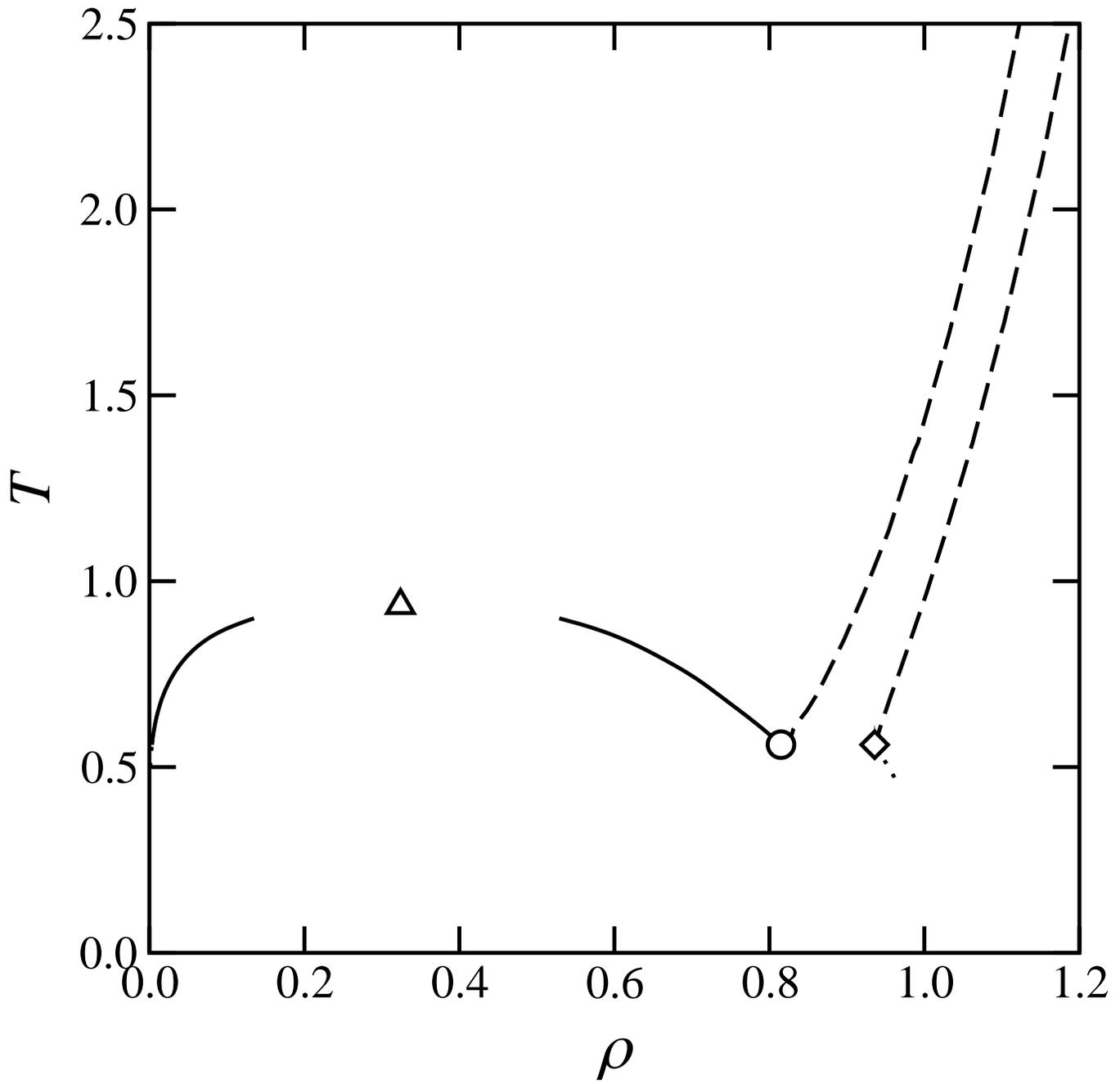

Figure 1a - Errington, Debenedetti and Torquato

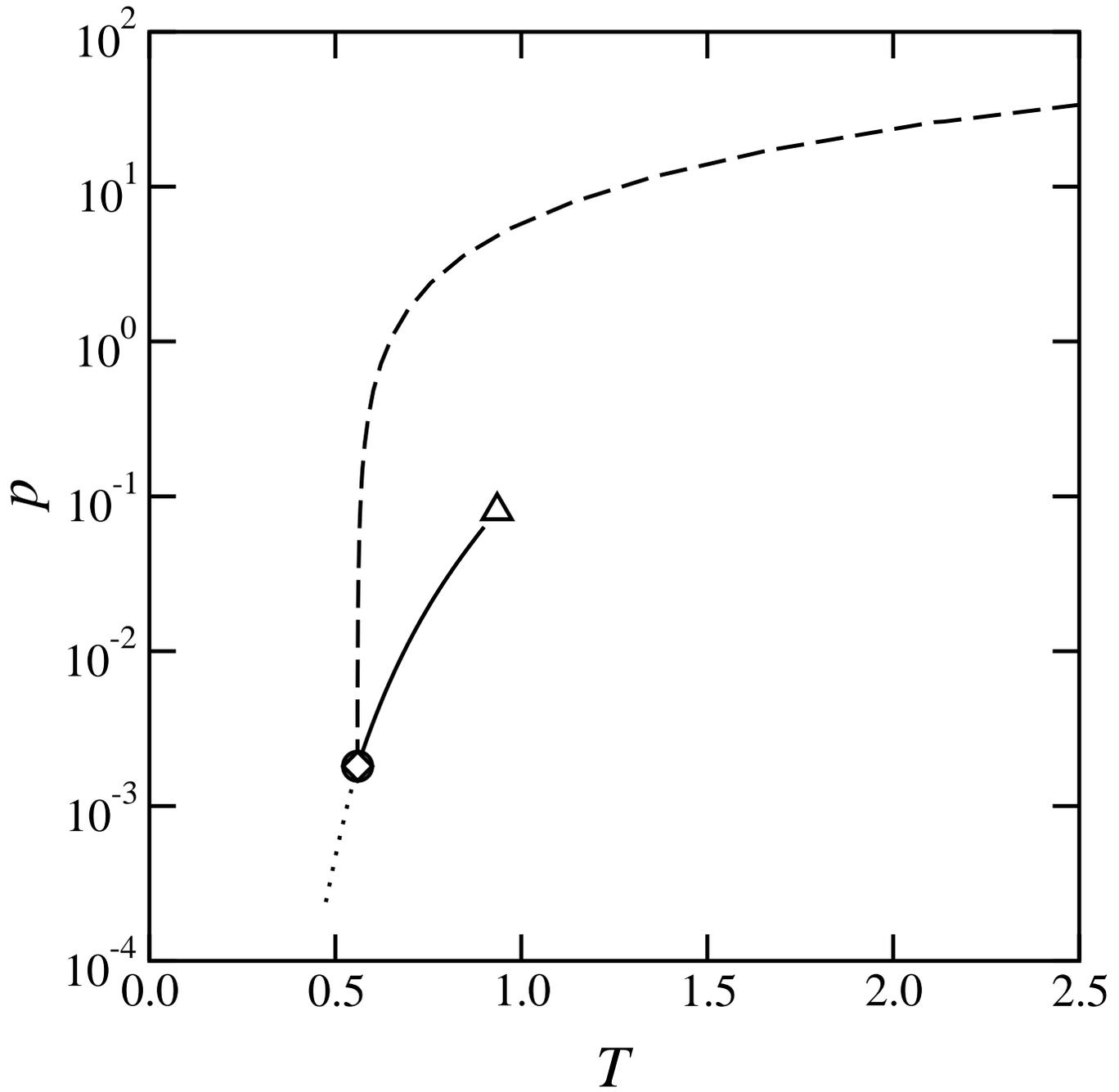

Figure 1b - Errington, Debenedetti and Torquato

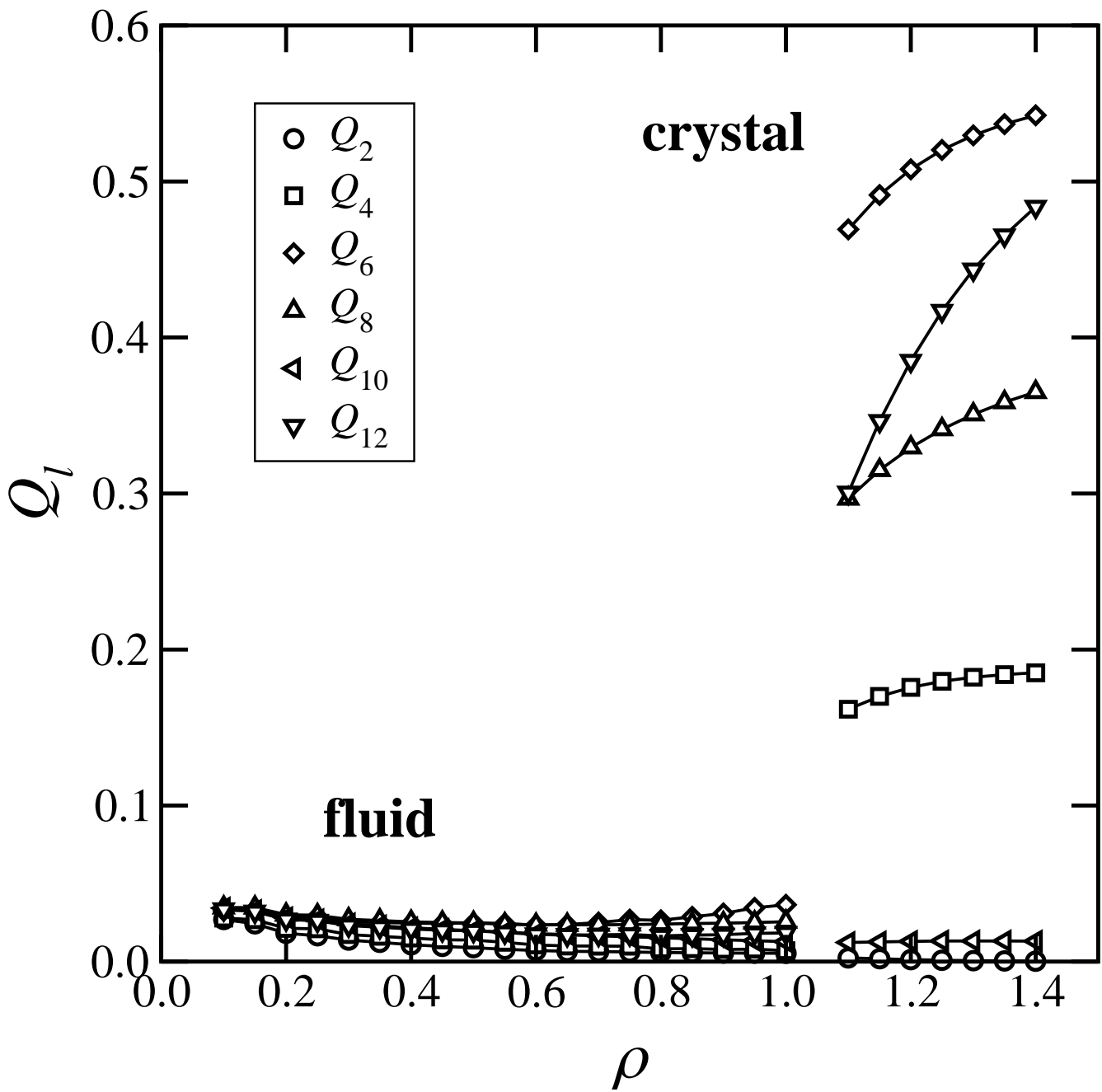

Figure 2 - Errington, Debenedetti and Torquato

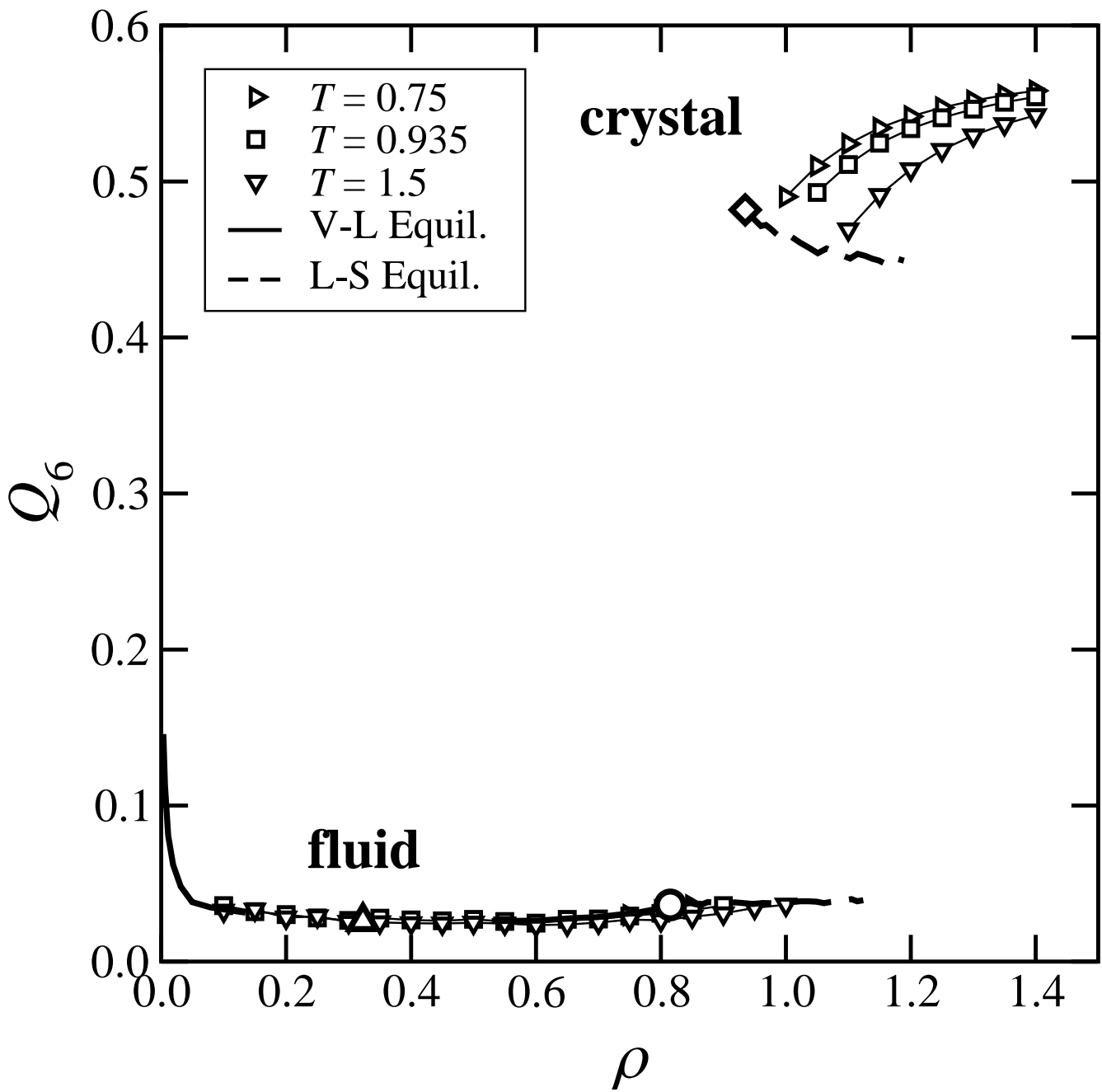

Figure 3 - Errington, Debenedetti, and Torquato

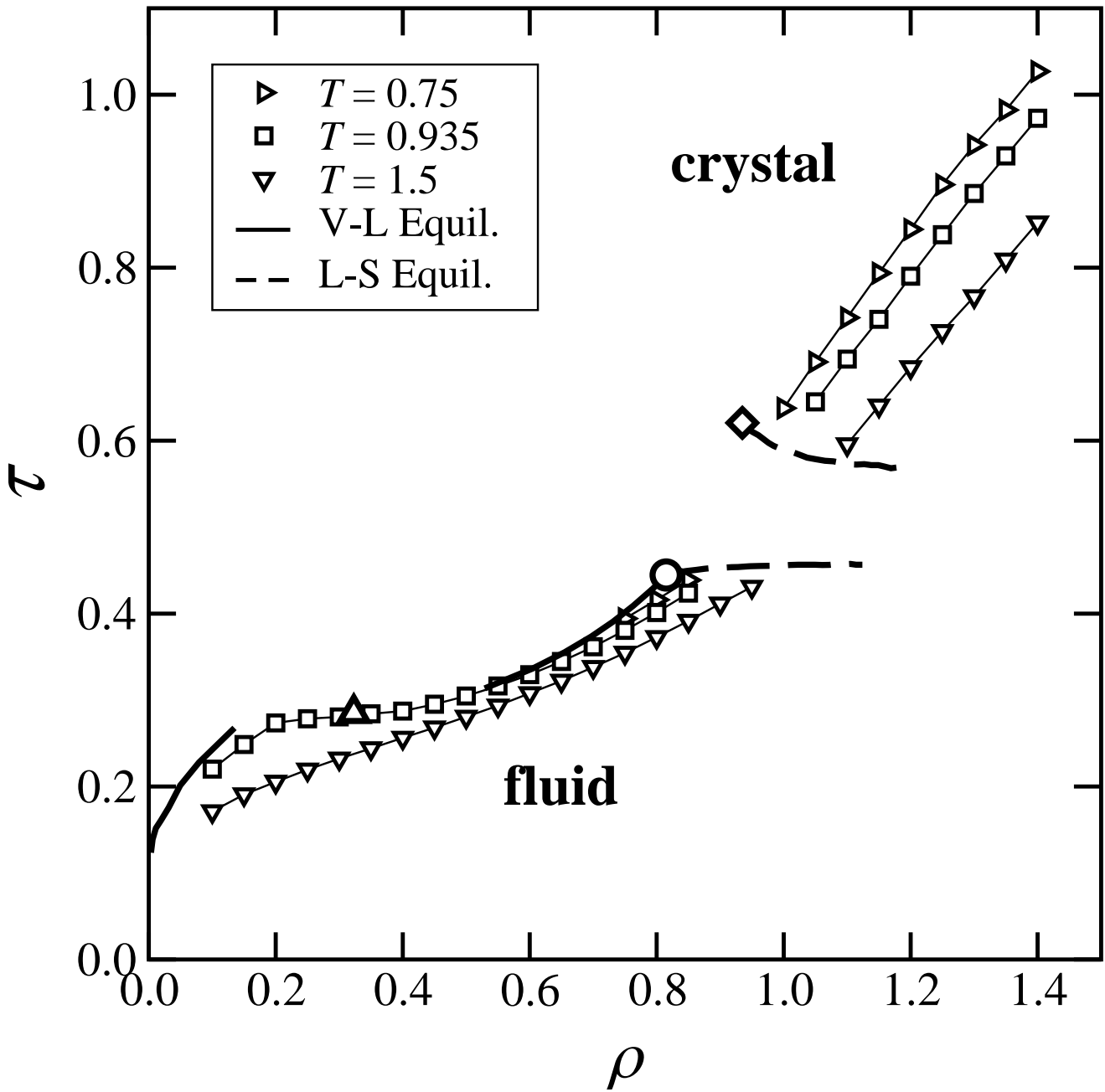

Figure 4 - Errington, Debenedetti and Torquato

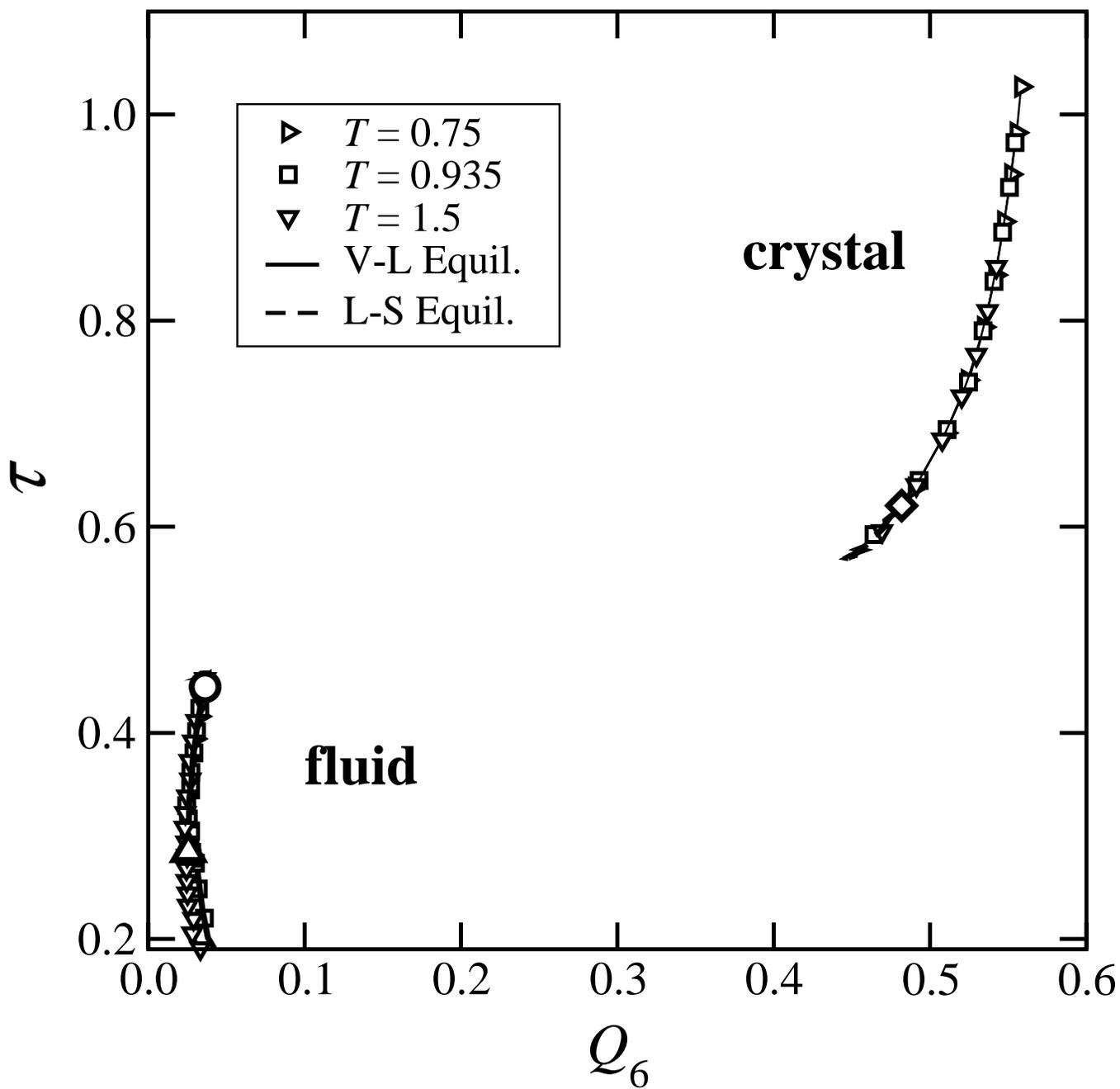

Figure 5 - Errington, Debenedetti and Torquato

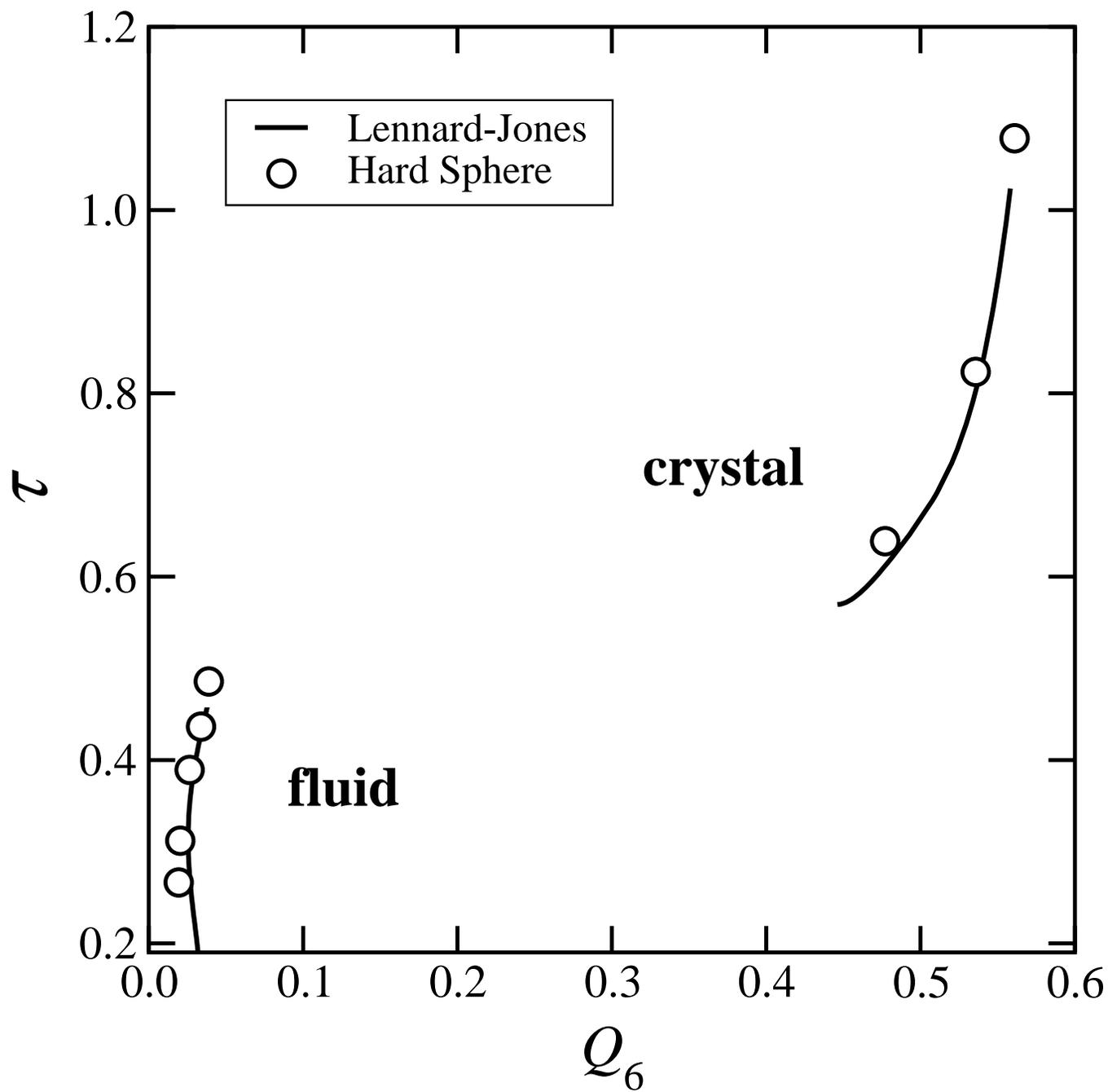

Figure 6 - Errington, Debenedetti and Torquato

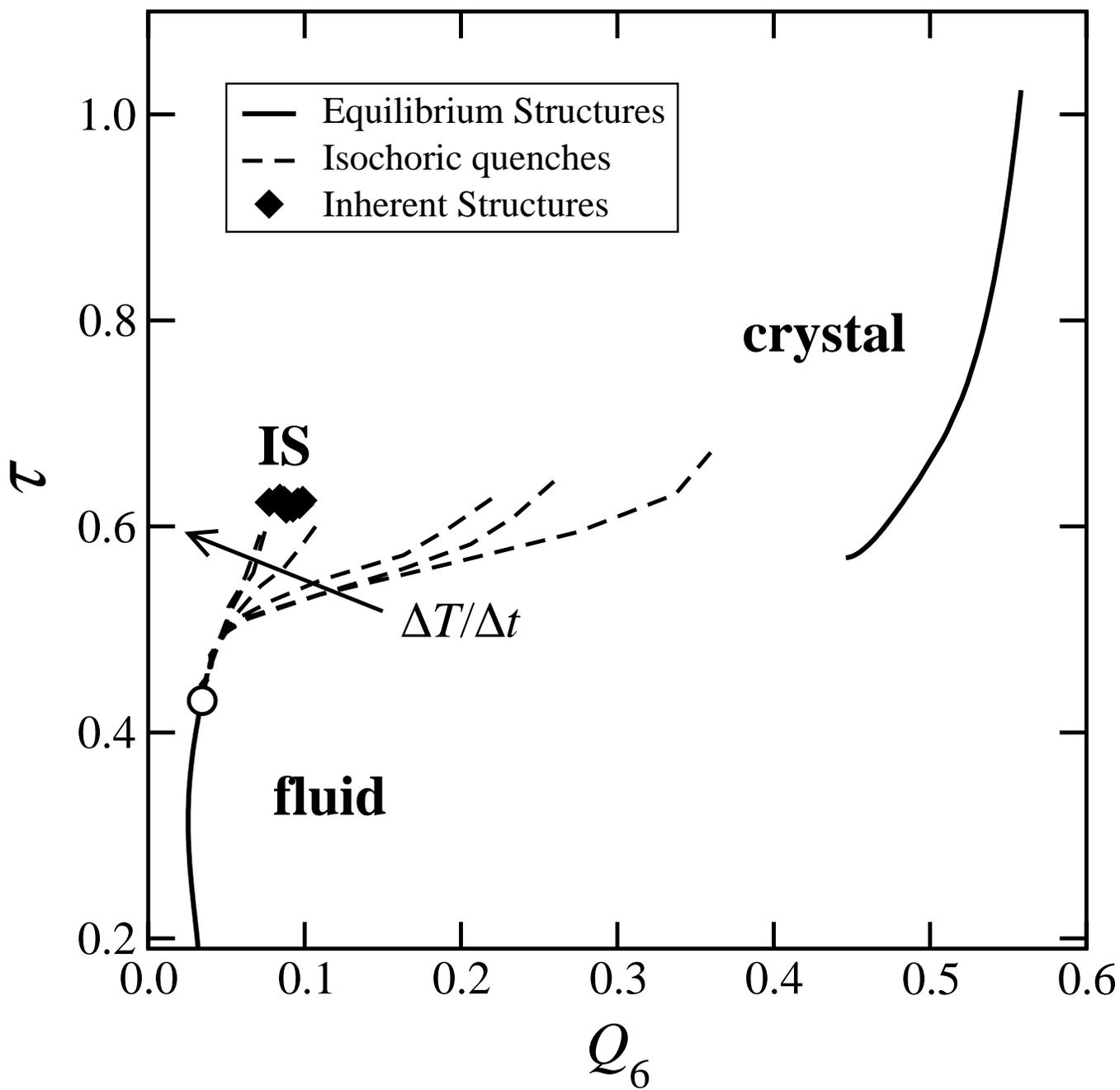

Figure 7 - Errington, Debenedetti and Torquato

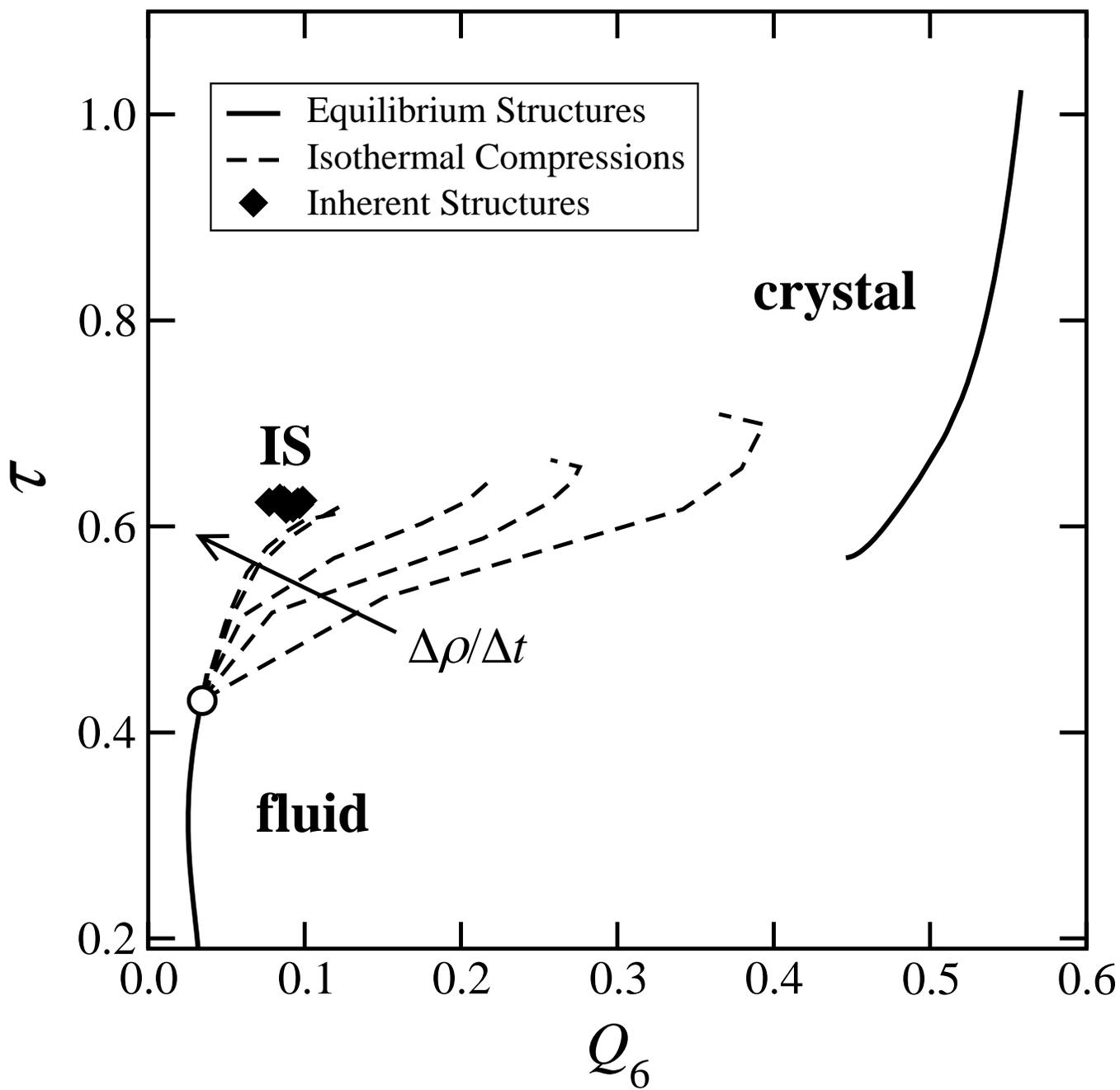

Figure 8 - Errington, Debenedetti and Torquato